\documentstyle[aps,twocolumn,psfig,floats]{revtex}

\draft 

\begin{document}
\twocolumn[\hsize\textwidth\columnwidth\hsize\csname
@twocolumnfalse\endcsname
\title{Energy transduction in periodically driven non-Hermitian systems}
\author{T. Alarc\'on, A. P\'erez-Madrid and J. M. Rub\'{\i}}
\address{Departament de F\'{\i }sica Fonamental and CER F\'{\i}sica de Sistemes Complexos,\\
Facultat de F\'{\i}sica,
Universitat de Barcelona,\\
Diagonal 647, 08028 Barcelona, Spain
}
\maketitle

\begin{abstract}
We show a new mechanism to extract energy from non-equilibrium fluctuations typical of periodically driven non-Hermitian systems. The transduction of energy between the driving force and the system is revealed by an \emph{anomalous} behavior of the susceptibility, leading to a diminution of the dissipated power and consequently to an improvement of the transport properties. The general framework is illustrated by the analysis of some relevant cases.
\end{abstract}


]
In the last years a growing interest in mechanisms for energy transduction
by rectification of unbiased thermal fluctuations has arose, partly motivated
by problems from cell biology. Several phenomenological models, categorized as thermal ratchets or Brownian motors,
have been proposed. These engines operate at molecular level \cite{kn:magnasco,kn:astumian,kn:prost,kn:hanggi}, although their potential implementation in a larger scale would be of evident interest. In this context, a number of methods for particle separation have been recently proposed based on several variants of the ratchet concept \cite{kn:prostnat,kn:drrat}. Motivated by this interest, we investigate how to take advantage of non-equilibrium fluctuations to optimize the energy consumption. We propose a new mechanism for energy transduction in a class of non-equilibrium systems when they are acted upon by a weak periodic force. The coupling between the external driving and the out-of-equilibrium fluctuations leads to a minimization of the dissipated power. The minimum occurs when the frequency of the external force matches a characteristic frequency of the system, thus manifesting a resonant behavior.

Let us consider the class of differential equations

\begin{equation}
\partial _{t}\Psi (\vec{x} ,t)\,=\,({\cal L}_{0}+\lambda (t){\cal L}%
_{1})\Psi (\vec{x} ,t),  \label{eq:1}
\end{equation}

\noindent describing the dynamics of a probability density or of a hydrodynamic field $\Psi (\vec{x} ,t)$, where $\vec{x}$ represents a coordinate. The dynamics is governed by the non-Hermitian operator ${\cal L}_{0}$ and is influenced by
the action of a periodic force which introduces the perturbation $\lambda(t){\cal L}_{1}$, with $\lambda(t)=\lambda_{0}e^{i\omega t}$. 

We will analyze the  response of the system to the external perturbation by using the standard procedure of the linear response theory.
Accordingly, Eq. (\ref{eq:1}) can be solved formally yielding

\begin{eqnarray}  \label{eq:2}
\Psi(\vec{x} ,t)&=&e^{(t-t_{0}){\cal L}_{0}}\Psi _{0}(\vec{x} ) 
\nonumber \\
&+&\int_{t_{0}}^{t}d\tau \,\lambda (\tau )e^{(t-\tau ){\cal L}_{0}}({\cal L}%
_{1}e^{(\tau -t_{0}){\cal L}_{0}}\Psi _{0}(\vec{x} )) \\
&\equiv& \Psi ^{(0)}(\vec{x} ,t)+\Delta \Psi (\vec{x} ,t), 
\nonumber
\end{eqnarray}

\noindent where $\Psi _{0}(\vec{x} )$ is the initial condition, which corresponds to the stationary state of Eq. (\ref{eq:1}) in the absence of the external force \cite{kn:risken,kn:ht}. Expansion of the term ${\cal L}_{1}e^{t{\cal L}%
_{0}}\Psi _{0}(\vec{x} )$ in series of the eigenfunctions of the
operator ${\cal L}_{0}$, {$\phi _{n}$} with eigenvalue $a_{n}$; $n=0,1,\,...$

\begin{eqnarray}
{\cal L}_{1}e^{t{\cal L}_{0}}\Psi _{0}(\vec{x} )=\sum_{n=0}^{\infty
}\{c_{n}\phi _{n}(\vec{x} )
+c_{n}^{*}\phi _{n}^{*}(\vec{x} )\},
\label{eq:3}
\end{eqnarray}

\noindent where $c_{n}$ are the corresponding coefficients in this expansion, then leads to 

\begin{equation}
\Delta \vec{x} (t)\,=\,\int \,\vec{x} \Delta \Psi (\vec{x}
,t)d\vec{x} \,=\,\int_{t_{0}}^{t}d\tau \vec{\chi} (t-\tau )\lambda (\tau ),  \label{eq:4}
\end{equation}

\noindent which defines the susceptibility $\vec{\chi} (t)$. 

We will assume the existence of a dominant time scale governing the
relaxation process, which corresponds to the $n=1$ mode in the expansion (\ref{eq:3}). Since the remaining modes decay faster we can truncate the
series retaining only the first term. Thus, considering only contributions of the first mode,

\begin{equation}
\vec{\chi} (t)\,=\,\vec{A}e^{a_{1}t}+\vec{A}^{*}e^{a_{1}^{*}t},  \label{eq:5}
\end{equation}

\noindent with $\vec{A}$ defined as

\begin{equation}
\vec{A}=c_{1}\int \vec{x} \phi
_{1}(\vec{x} )d\vec{x},  \label{eq:6}
\end{equation}

\noindent the explicit expression of the susceptibility as a function of
the frequency follows from the Fourier transform of Eq. (\ref{eq:5})

\begin{equation}  \label{eq:7}
\vec{\chi}(\omega )\,=\frac{\vec{A}}{I_{1}}\frac{1}{\beta -i(\alpha +1)}+\frac{\vec{A}^{*}}{I_{1}}\,\frac{1}{\beta-i(\alpha -1)}
\end{equation}

\noindent where $a_{1}\equiv R_{1}+iI_{1}$, $\beta \equiv R_{1}/I_{1}$ and $%
\alpha \equiv \omega /I_{1}$, with $R_{1}$ and $I_{1}$ being the real and imaginary parts of $a_{1}$, respectively

In Fig. 1, we have plotted the modulus of the susceptibility as
a function of $\alpha$ for different values of $\beta$. During the
relaxation process of non-equilibrium fluctuations, the susceptibility
undergoes a resonant behavior when the frequency of the force matches the imaginary part of
the first eigenvalue of the non-perturbed operator ${\cal L}_{0}$. This behavior reveals the resonant coupling between the periodic force and the
non-equilibrium source, responsible for the non-Hermitian nature of ${\cal L}%
_{0}$. 

The appearance of this resonance leads to a diminution of the dissipation in the relaxation process of the fluctuations of $\vec{x}$. To illustrate our assertion, let us suppose that Eq. (\ref{eq:1}) represents the non-Hermitian Fokker-Planck equation

\begin{eqnarray}\label{eq:8}
\partial _{t}\Psi (\vec{x} ,t)\, =\,-\nabla_{\vec{x}}\cdot(\vec{v}(
\vec{x})\Psi-D\nabla_{\vec{x}}\Psi-b\lambda(t)\Psi\nabla_{\vec{x}}U(\vec{x})),
\end{eqnarray}

\noindent where $\vec{v}(\vec{x})$ is a non-potential drift, $b$ a mobility, $D=K_{B}Tb$ the corresponding diffusion coefficient and $U(\vec{x})$ the potential related to the external force. In this case the operators ${\cal L}_{0}$ and ${\cal L}_{1}$ act on the field as

\begin{eqnarray}\label{eq:9}
\nonumber {\cal L}_{0}\Psi&=&-\nabla_{\vec{x}}\cdot(\vec{v}(
\vec{x})\Psi-D\nabla_{\vec{x}}\Psi),\\
{\cal L}_{1}\Psi&=&\nabla_{\vec{x}}\cdot(b\Psi\nabla_{\vec{x}}U(\vec{x})).
\end{eqnarray}

Among the physical realizations of the model described by Eq. (\ref{eq:8}) we can quote the case of a Brownian particle advected by a constant drift $\vec{v}$ acted upon by a force $\vec{F}(\vec{x},t)$, or a field-responsive particle in a vortex flow under the influence of an oscillating magnetic field \cite{kn:bacri,kn:vorticity}.  

We are interested in analyzing the energy dissipated by the system in the dynamic process governed by Eq. (\ref{eq:8}). The dissipated power is 

\begin{equation}\label{eq:10}
P=-\int\,d\vec{x}\vec{J}\cdot\nabla_{\vec{x}}\mu,
\end{equation}

\noindent where $\vec{J}=-D\nabla_{\vec{x}}\Psi-b\lambda(t)\Psi\nabla_{\vec{x}}U(\vec{x})$ is the diffusion current and $\mu=K_{B}T\ln\Psi(\vec{x},t)+\lambda(t)U(\vec{x})$ the corresponding chemical potential. It splits up into two contributions, one coming from the underlying diffusion process and another accounting for the power supplied by the external force. Since we are interested in the effects of the external field, we focus on the externally supplied power, 

\begin{eqnarray}\label{eq:11}
P_{F}&=&-\lambda(t)\int\,d\vec{x}\vec{J}\cdot\nabla_{\vec{x}}U=\int\,d\vec{x}\vec{J}\cdot\vec{F}(\vec{x},t).
\end{eqnarray}

\noindent By expanding $\vec{F}(\vec{x},t)$ in series of the eigenfunctions of ${\cal L}_{0}$

\begin{eqnarray}\label{eq:12}
\vec{F}(\vec{x},t)=\vec{F}_{0}(t)+\sum_{n\neq 0}(\vec{F}_{n}(t)\phi_{n}(\vec{x})+\vec{F}_{n}^{*}(t)\phi_{n}^{*}(\vec{x})),
\end{eqnarray}

\noindent and substituting this equation into Eq. (\ref{eq:11}) we achieve

\begin{eqnarray}\label{eq:13} P_{F}=\vec{F}_{0}(t)\cdot\left\{\frac{d\langle\vec{x}\rangle}{dt}-\langle\vec{v}(\vec{x})\rangle\right\}+\sum_{n\neq 0}2\Omega{\sf Re}\{\vec{F}_{n}(t)\cdot\vec{J}_{n}^{*}\},
\end{eqnarray}

\noindent where $\Omega$ is the volume of the system. To obtain Eq. (\ref{eq:13}) we have used the result 

\begin{equation}\label{eq:14}
\int d\vec{x}\,\vec{J}=\frac{d\langle\vec{x}\rangle}{dt}-\langle\vec{v}(\vec{x})\rangle.
\end{equation}

\noindent This expression follows from the Fokker-Planck Eq. (\ref{eq:8}) through the definition of $\vec{J}$.

The quantity of interest is the dissipated power averaged over the period of the external force, 

\begin{equation}  \label{eq:16}
\overline{P}(\omega)\,=\,\frac{\omega}{2\pi}\int_{0}^{2\pi/\omega}dt\,P_{F}.
\end{equation}

For the particular case of the Brownian particle, $P_{F}$ reads

\begin{eqnarray}\label{eq:15}
P_{F}=\vec{F}_{0}(t)\cdot\frac{d\langle\vec{x}\rangle}{dt}+\sum_{n\neq 0}2\Omega{\sf Re}\{\vec{F}_{n}(t)\cdot\vec{J}_{n}^{*}\}.
\end{eqnarray}

\noindent In Fig. 2a we have plotted the corresponding $\overline{P}(\omega)$. To this end we have assumed that the motion of the particle takes place in two dimensions, with periodic boundary conditions, thus, the set of eigenfunctions $\phi_{\vec{k}}(\vec{x})$ are the Fourier modes with eigenvalues $a_{\vec{k}}=-Dk^{2}-i\vec{v}\cdot\vec{k}$. We have considered a force of the form $\vec{F}(\vec{x},t)=\lambda(t)(1+\cos(\vec{k}\cdot\vec{x}))\hat{v}$, where $\hat{v}$ is the unit vector pointing along the direction of the drift. The figure shows that $\overline{P}(\omega)$ achieves its minimum value at the resonant frequency. The negative character of this quantity indicates that the system is acting as a generator.

In Fig. 2b we have represented that quantity for the more complex case of the mesoscopic dynamics of a field-responsive Brownian particle in a vortex flow with vorticity $\vec{\omega}_{0}$, under an oscillating magnetic field, $\vec{H}(t)$, \cite{kn:bacri,kn:vorticity}. In this case, the power supplied by the external force is

\begin{equation}\label{eq:17}
P_{F}=\vec{H}(t)\cdot\left\{\frac{d\langle\vec{M}\rangle}{dt}-\vec{\omega}_{0}\times\langle\vec{M}\rangle\right\},
\end{equation}

\noindent where $\vec{M}$ is the magnetization. This figure shows that in this case the transduction of energy occurs in two different regimes. In the low frequency regime $\overline{P}_{0}$, the power dissipated in the Debye relaxation \cite{kn:degroot}, achieves negative values while $\overline{P}_{c}$, the energy dissipated due to the coupling between the drift and the external force, takes negative values for high frequencies. In both cases the averaged dissipated power $\overline{P}(\omega)$ exhibits a minimum value at the resonant frequency, showing the resonant character of the mechanism for energy transduction.

The analysis of the energy dissipation allows us to study the transport properties. The presence of an external driving, forces the system to move with a mean velocity, $\vec{v}_{m}$, different from the drift $\vec{v}(\vec{x})$, thus leading to the appearance of a drag force, $\vec{F}_{d}=-\vec{\vec{\kappa}}\cdot\vec{v}_{m}$  \cite{kn:ll}, with $\vec{\vec{\kappa}}$ a friction tensor accounting for dissipation in the system. The resulting dissipated power is

\begin{equation}\label{eq:18}
\overline{P}=\vec{v}_{m}\cdot\vec{\vec{\kappa}}\cdot\vec{v}_{m}.
\end{equation}

In the case of the Brownian particle advected by a constant drift, the friction tensor is a scalar. Its expression follows form Eq. (\ref{eq:18})

\begin{eqnarray}\label{eq:20} \kappa_{B}=\frac{1}{v_{m}^{2}}\overline{P},
\end{eqnarray}

\noindent whose behavior is essentially shown in Fig. 2a.  

For the field-responsive particle the dissipated power given through Eq. (\ref{eq:17}) consists of two independent contributions corresponding to longitudinal ($\overline{P}_{0}$) and transversal ($\overline{P}_{c}$) effects, with respect to the direction of the magnetic field. Consequently, associated to the last one, which corresponds to the viscous dissipation occurring when the magnetic field acts on the fluid, we can define a friction coefficient as

\begin{equation}\label{eq:21}
\kappa_{F}=-\frac{1}{(v_{m}^{T})^{2}}\overline{\vec{\omega}_{0}\cdot(\langle\vec{M}\rangle\times\vec{H})}.
\end{equation}  

\noindent where $v_{m}^{T}=\omega_{0}a$ is the transversal component of $\vec{v}_{m}$, with $a$ the radius of the particle and $\omega_{0}$ the modulus of the vorticity. 

It is interesting to analyze the behavior of this quantity in terms of the parameter $\beta$, which is given in this case by $D_{r}/\omega_{0}$, with $D_{r}$, the rotational diffusion coefficient. As can be seen in Fig. 3, as $\beta$ grows the friction coefficient $\kappa_{F}$ becomes positive in the entire frequency range. In fact the frequency for
which $\kappa_{F}=0$ goes to infinity in the limit of a Hermitian dynamics. Fig.
1 shows something analogous, that is, the resonance disappears when $\beta$
grows. Thus, we conclude that the phenomenon described is genuine of non-Hermitian systems.

The anomalous behavior exhibited by the friction coefficient is a direct consequence of the form of the susceptibility. In equilibrium, the fluctuation-dissipation theorem, implying that $\omega{\sf Im}\chi_{x}(\omega)\geq 0$, manifests that during the relaxation of the fluctuations around an equilibrium state the system always dissipates energy. Nonetheless, in the non-equilibrium case shown in Fig. 4 the imaginary part of the susceptibility achieves negative values for positive frequencies thus violating the aforementioned inequality. In this figure, obtained for the particular case of the Brownian particle in a constant drift, the analytical results are compared with numerical results from the corresponding Langevin equation by means of a second order Runge-Kutta method \cite{kn:simul}. For the field-responsive particle, we obtain the same behavior, which essentially corresponds to $\overline{P}_{0}$, plotted in Fig. 2b. This fact indicates that the system is generating energy instead of dissipating power, which manifests at a macroscopic level through the diminution of the friction coefficient. An  anomalous behavior of the response was first discussed in the context of current generation by noise induced symmetry-breaking in coupled Brownian motors \cite{kn:rkvdh}. In addition, several similar phenomena have been reported in \cite{kn:reim,kn:buceta}. These papers were focused on specific models, raising the question under which circumstances this behavior arises. In this letter, we have found the conditions for these phenomena to occur in a quite general class of systems.

Notice that our results differ from the ones obtained when LRT is applied to non-equilibrium Hermitian systems. For example, in Ref. \cite{kn:ht} fluctuation-dissipation-type relationships are derived, by assuming that the relaxation occurs as in the case of an equilibrium state. In the present context, this assumption does not hold, since we are dealing with non-Hermitian systems whose eigenvalue spectrum is complex. Consequently, perturbations do not relax exponentially as they do in equilibrium.

The theoretical framework discussed in this paper can be applied to a number
of problems formulated in terms of non-Hermitian dynamics. Among them, one
could mention the transport of classical particles advected by a quenched%
\cite{kn:wang} velocity field. This process, governed by a Fokker-Planck
equation with random drift, models the diffusion in porous media. It
has been recently shown \cite{kn:wang} that, when the velocity field
displays correlations in both longitudinal and transversal directions,
the eigenvalues occupy a finite area in the complex plane. Consequently, this system evolves according to a non-Hermitian dynamics. If the particles
can respond to an external field the same phenomenology described in this
paper holds, i.e., as a consequence of the
diminution of the dissipation transport through the porous medium becomes enhanced. The phenomenon we study may also arise in population biology
problems, in particular in the generalization of the Malthus-Verhulst growth
model proposed by Nelson and Schnerb \cite{kn:nelson2}. The linearization of
this model around its steady state yields a non-Hermitian evolution equation. When a periodic driving is introduced as a time
dependence of the resources of the medium, the minimum achieved by the dissipated
energy is now related to a resonant optimization of these
resources.

Additionally, our approach might unify the explanation of other resonant
transport phenomena previously reported. The
Senftleben-Beenakker effect observed in gases of polyatomic molecules shares the  phenomenology inherent to our model. This effect occurs when the gas is under the action of both a constant magnetic field and an oscillating field parallel to the first one \cite{kn:beenakker}. Larmor precession causes the
non-Hermiticity of the Boltzmann equation describing the
dynamics of this system. The resonant frequency is related to Larmor's frequency. Under these conditions, the viscosity of the gas manifests a non-monotonous behavior as a function of the frequency similar to the one depicted in Fig. 3. Another example exhibiting analogous characteristics is the {\em negative viscosity}
effect observed in field-responsive fluids \cite{kn:bacri,kn:vorticity},
under a non-potential flow and submitted to an AC field. This effect consists of a diminution of the viscosity due to the presence of the periodic field. The field-responsive phase acts as a transmitter of energy between the external force and the system, leading to an improvement of the transport. 

In summary, we have proposed a mechanism for energy transduction in non-equilibrium systems, based on the possibility of extracting energy from the relaxation process of out-of-equilibrium fluctuations. Under the action of an oscillating force, systems which evolve according to non-Hermitian dynamics act as transducers. Consequently, the energy dissipated in the system diminishes achieving its minimum value when the frequency of the external driving matches the resonant frequency. This diminution of the amount of energy dissipated has a strong influence on the macroscopic properties of the system, leading to an enhancement of the transport or more generally to an optimization of the consumption of energy. Due to the intrinsic non-equilibrium nature of the fluctuations,  energy transduction does not require any further ingredient, as occurs in ratchet-like engines, in which the presence of a parity-symmetry-breaking potential is an unavoidable condition for transferring energy.       

The authors thank David Reguera for helpful discussions. This work has been
supported by DGICYT of the Spanish Government under grant PB98-1258. One of
us (T. Alarc\'{o}n) wishes to thank to DGICYT of the Spanish Government for
financial support.




\begin{figure}[htb]
\centerline{\psfig{file=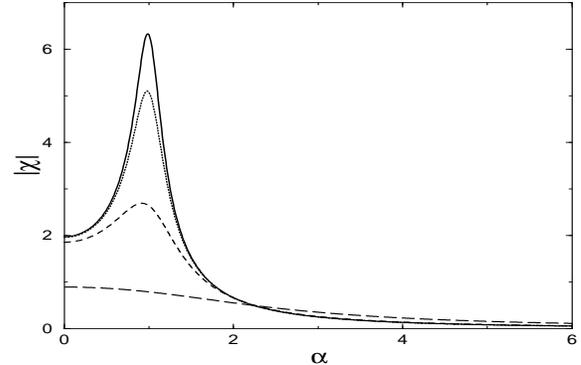,height=5cm,width=7.5cm}}
\caption{Non-dimensional modulus of the susceptibility as a function of the parameter $\alpha$. Continuous line corresponds to $\beta=0.1$, dotted line
to $\beta=0.5$ and dashed line to $\beta=1$. The resonance fades away practically for $\beta\approx 10.$}
\label{fig:1}
\end{figure}

\begin{figure}[htb]
\centerline{\psfig{file=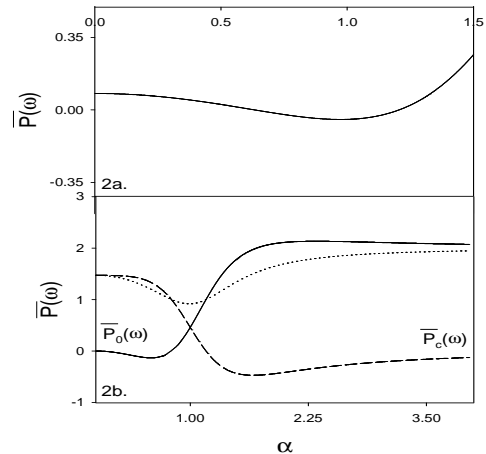,height=9cm,width=8cm}}
\caption{2a.-Non-dimensional dissipated power for a particle under a constant drift as a function of the
non-dimensional parameter $\alpha$.2b.-Same for a field-responsive fluid. The dotted line corresponds to the total
dissipation $\overline{P}(\omega)$.}
\label{fig:2}
\end{figure}

\begin{figure}[htb]
\centerline{\psfig{file=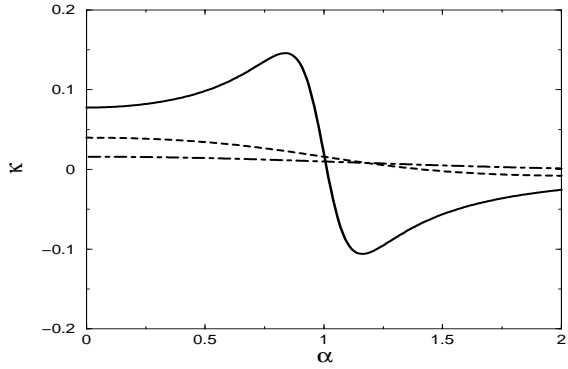,height=5cm,width=7.5cm}}
\caption{Non-dimensional friction coefficient as a function of the parameter $\alpha$ for different values of $\beta$. Continuous line corresponds to $\beta=0.1$, dotted line to $\beta=0.5$ and dashed line to $\beta=1$.}
\label{fig:3}
\end{figure}

\begin{figure}[htb]
\centerline{\psfig{file=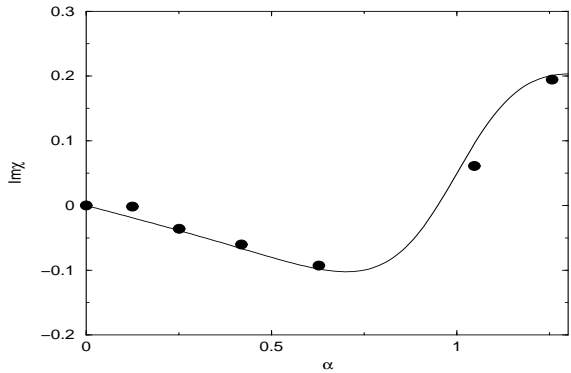,height=5cm,width=7.5cm}}
\caption{Non-dimensional imaginary part of the susceptibility for a Brownian particle advected by a constant drift as a function of the parameter $\alpha$. Solid line corresponds to the analytical computation whereas dots have
been obtained from numerical simulations.}
\label{fig:4}
\end{figure}

\end{document}